\documentclass[aip, reprint]{revtex4-1}
\usepackage{amsmath}    % need for subequations
\usepackage{graphicx}   % need for figures
\usepackage{textcomp}

\begin{document}

\title{Order-Disorder Behavior at Thin Film Oxide Interfaces}

\author{Steven R. Spurgeon}
\affiliation{Energy and Environment Directorate \\ Pacific Northwest National Laboratory \\ Richland, Washington 99352}
\email{steven.spurgeon@pnnl.gov}

\date{\today}

\begin{abstract}

Order-disorder processes fundamentally determine the structure and properties of many important oxide systems for energy and computing applications. While these processes have been intensively studied in bulk materials, they are less investigated and understood for nanostructured oxides in highly non-equilibrium conditions. These systems can now be realized through a range of deposition techniques and probed at exceptional spatial and chemical resolution, leading to a greater focus on interface dynamics. Here we survey a selection of recent studies of order-disorder behavior at thin film oxide interfaces, with a particular emphasis on the emergence of order during synthesis and disorder in extreme irradiation environments. We summarize key trends and identify directions for future study in this growing research area.

\end{abstract}


\maketitle

\section{Introduction}

Understanding and controlling order in metals\cite{George2019} and ceramics\cite{Oses2020} represents a grand materials science challenge with broad implications for transformative technologies, including quantum computing, electronics, and clean energy. Order-disorder processes govern important functionalities, such as electronic and magnetic order, that are intimately linked to the dynamics of atomic-scale structure, chemistry, and defects. Many experimental and theoretical frameworks have been developed to understand order-disorder behavior in different materials classes, including metals, ceramics, and polymers.\cite{w1976solid} However, most frameworks have focused on the behavior of bulk materials, engineered through traditional solidification and solid state processing routes. The emergence of powerful, atomically-precise synthesis techniques for semiconductors and oxides in the late twentieth century introduced a new class of nanostructured materials, which are grown far from equilibrium and are not well described by conventional bulk thermodynamics.\cite{Chambers2010b} Thin film materials exhibit novel and desirable properties distinct from their bulk counterparts,\cite{Huang2018} such as emergent ferroic ordering, enhanced ion conduction, and unique radiation responses, which has motivated efforts to engineer their functionality through intricately designed synthesis routes.\cite{MacManus-Driscoll2020, Mundy2016c}

Despite extensive past work, the field is still limited by an incomplete understanding of growth pathways and defect formation mechanisms, which lead to deviations from idealized crystal structures and poorly controlled order-disorder behavior. The two most common classes of functional oxides---perovskites and pyrochlores---possess flexible crystal structures that can host a range of multivalent cation species, lattice defects, and reconfigurations.\cite{Tuller2011,Chakoumakos1984,Subramanian1983} The strong coupling between lattice, spin, and charge degrees of freedom in these systems provides a handle to tune properties.\cite{Ismail-Beigi2017} Many prior thin film engineering studies have focused on the impact of substrate constraints on growth pathways and resulting interface structure and properties.\cite{Martin2017} More recently, efforts have turned to probing the dynamic nature of the thin film growth process resulting from the balance of thermodynamic and kinetic limitations.\cite{Smith2017} The field has also become increasingly concerned with the influence of extreme conditions\cite{Shamblin2016a,Swallow2017} on interface and phase stability. It is becoming clear that successful predictive design of oxide interfaces requires a more nuanced grasp of the factors mediating order-disorder behavior.

Recent advances in characterization and modeling tailored to probing interface behavior have begun to provide new insight into order-disorder processes. The development of \textit{in situ} X-ray diffraction (XRD) methods during deposition, for example, has revealed complex structural rearrangements during growth.\cite{Cook2019,Andersen2018,Chang2016,Bein2015,Lee2014} Area-averaged scattering methods can be complemented by aberration-corrected scanning transmission electron microscopy (STEM) and electron energy loss spectroscopy (STEM-EELS) techniques, which can investigate localized, aperiodic perturbations, defects, and chemical states of phase boundaries.\cite{Guo2020, Spurgeon2017a} These experiments are underpinned by a theoretical framework of \textit{ab initio} simulations, which can be linked to scattering and microscopy measurements. A variety of codes\cite{Ophus2017,Allen2015} have been developed for image and spectral simulations based on relaxed theory models, allowing defect configurations to be extracted with a high level of accuracy. 

However, many questions remain unanswered about the exact nature of order-disorder behavior at thin film oxide interfaces, even in common perovskite and pyrochlores. It is presently still difficult to adequately characterize growth dynamics of many oxide systems with sufficient spatial, chemical, and temporal resolution. In turn, we are unable to predict optimal synthesis pathways and defects with a high degree of accuracy,\cite{Lloyd2017} precluding us from achieving targeted structures and controlling their evolution in extreme, non-equilibrium environments. Similarly, disordering behavior of model oxide interfaces has not been widely studied, leading to an incomplete picture of defect formation in specific, well-defined interface configurations. The roles of the interface versus the bulk in defect generation and transport are also currently unclear, motivating efforts to examine this behavior in greater detail through correlative experiment and theory. In this critical review, we focus on more recent literature and progress in this area. Our aim is not to be exhaustive, but rather to highlight a selection of ongoing developments to inspire future research efforts in this fast evolving field.

\section{Origins of Order-Disorder Behavior}

Order-disorder phenomena have been intensely studied since before the advent of modern materials science and solid state physics.\cite{w1976solid} These processes broadly describe the transition of material from a crystalline to amorphous state and vice versa, with varying degrees of order in between. The nature of this transition and its mediation by a hierarchy of point, cluster, line, and extended defects is a richly complex topic with enormous implications for properties of materials. For the purposes of this review, we will consider defect types specific to interfaces of perovskite and pyrochlore oxides. These materials are ubiquitous components of heterostructures, possessing high structural and chemical compatibility as well as desirable properties. To better understand defect behavior and its subsequent evolution, particularly in non-equilibrium synthesis and processing conditions, we first discuss the lattice and energetic drivers that determine the stability of these components and their interfaces.

\subsection{Bulk Perovskites and Pyrochlores}

The ideal $AB$O$_3$ cubic perovskite structure possesses a $Pm\bar{3}m$ space group consisting of cubic close-packed $A$O$_3$ octahedra and corner-sharing $B$O$_3$ octahedra, where $A$ and $B$ are transition metal and rare-earth cation species such as Fe, Ti, Sr, La, and others. Deviations from the ideal cubic structure are described by the Goldschmidt tolerance factor ($r_A / r_B$),\cite{Goldschmidt1926} which relates the geometric compatibility of $A$--O and $B$--O bond lengths and resulting lattice distortions. For tolerance factors $<1$ the coordination number of the $A$-site cation species can be lowered through correlated $B$O$_6$ octahedral tilting, which has been described by Glazer \textit{et al.},\cite{Glazer1972} as well as Jahn-Teller distortions of the cation species. Alternatively, for tolerance factors $>1$, the $A$O$_3$ octahedra can adopt partial or full hexagonal stacking sequences, often with the formation of extensive $B$-site cation vacancies. Order in the perovskite system can occur on both cation sublattices, as well on anion sites. Differences in preferred coordination, valence, and size of cation species can lead to $A$-site ordering (usually with cation vacancies) or $B$-site ordering between two metal cations (yielding the double perovskite structure),\cite{King2010} producing materials with novel dielectric and ferroic behavior.\cite{Davies2008} Alternatively, in non-stoichiometric perovskites, oxygen vacancies can order into elaborate networks, as in the case of brownmillerites and cuprate superconductors.\cite{Stolen2006} Because of the strong connection between vacancy formation and properties, efforts to probe and harness vacancy populations has attracted considerable attention.\cite{Gunkel2020}

The ideal $A_2B_2$O$_7$ pyrochlore structure is a derivative of the $Fd\bar{3}m$ fluorite structure, which possesses a face-centered cubic arrangement of the (usually-larger) $A$-site cation species and full tetrahedral interstitial occupancy of the $B$-site cation species. However, the pyrochlore derivative exhibits alternate ordering of the $A$- and $B$-site species in rows along the $<$110$>$ direction. Most often, 3+ valence $A$-sites are fully 8-fold oxygen coordinated, while 4+ valence $B$-sites are only 6-fold oxygen coordinated, with two vacant anion sites to maintain charge neutrality. The difference in cation size is a major factor in ordering of the structure and randomization of the cation species must be accompanied by simultaneous vacancy disorder in the anion sublattice.\cite{Wuensch2000} This unique coupling between cation and anion disorder imparts potentially useful ionic conductivity and electronic properties.\cite{Shlyakhtina2012} Interestingly, these materials have been shown to exhibit local fluctuations in order that can affect properties.\cite{Shamblin2016a} In addition, the alloying of different cation species provides a route to control the degree of ordering into the pyrochlore structure, as the cation antisite formation energy scales with chemical substitution.\cite{Sickafus2000} Because of the similarity between the ordered pyrochlore and disordered fluorite structure, the radiation effects community has examined pyrochlores in high-radiation environments, such as those encountered in nuclear waste forms and reactor components.\cite{Lang2010,Chartier2002,Sickafus2000}

\subsection{Thin Film Interfaces}

The development of refined layer-by-layer deposition techniques has revolutionized the design of high-purity oxide and semiconductor thin films. Here we focus primarily on the various modes of molecular beam epitaxy (MBE), including elemental, chemical, metal-organic, and hybrid variants, which are among the most precise methods to engineer film structure and chemistry.\cite{Prakash2019, Brahlek2018, Schlom2015} MBE enables epitaxial deposition of a wide variety of metal species in a clean, ultra-high vacuum environment under controlled oxygen partial pressure without the injection of energetic ion species. In a typical experiment, beams of metal atoms are generated produced by metal effusion cells, generating an atomic flux pointing toward a single crystalline substrate. An oxygen source (O, O$_2$, or O$_3$) is simultaneously introduced, ideally leading to uniform, fully oxidized atomic layers. \cite{Chambers2010b} Growth can be performed in co-evaporative or shuttered modes and monitored \textit{in situ} using reflection high-energy electron diffraction (RHEED) to achieve precise monolayer-level control of film configurations. MBE has been used to engineer a wide variety of III-V / II-VI semiconductors, superconductors, and ferroic oxides, giving rise to a whole field of interface engineering.\cite{Huang2018,Schlom2015}

There are many additional defect types that may form during nanostructuring of oxides, owing to the added mobility of alloying elements and energy costs associated with interface formation. In general, the most common defects are strain-induced structural distortions, misfit dislocation formation, cation intermixing, and oxygen vacancies.\cite{Du2014,Du2017,Spurgeon2016,Zhang2015} Oxides are typically grown on single-crystal substrates with similar lattice parameters to the desired film material. For thin layers on low lattice mismatch (a few \%) substrates, epitaxy can usually be achieved through coherent lattice strain and local bond distortions in the film. Because of strong electron-lattice coupling, this process can modify the material's electronic, magnetic, and optical properties in the vicinity of the interfaces, as has been well established for ferroic materials.\cite{Martin2017, Damodaran2016} Alternatively, for thicker films or those possessing greater lattice mismatch, misfit dislocations can form to relax the overall lattice strain, leading to semicoherent or incoherent interfaces.\cite{Uberuaga2019} The far-reaching strain field around these dislocations (several nm) can lead to the formation of intricate and complex networks, which in turn possess unique chemical, transport, and radiation responses.\cite{Pilania2020,Chen2019, Bagues2018, Shutthanandan2017, Martinez2016, Choudhury2014} In addition to these extended defects, more localized cation defects can also easily form in response to the interface strain and charge state, depending on the choice of substrate and its surface termination. The similar size of multivalent cation species, coupled with the elevated temperatures used during deposition, provide ample opportunity for cation intermixing, which can completely alter the character of the interface.\cite{Chambers2010} Finally, as is the case for bulk perovskites and pyrochlores,\cite{Ganduglia-Pirovano2007} oxygen defects can also easily form in the vicinity of interfaces in response to competing ionic, charge, and transport processes.\cite{Gunkel2020} These defects can exist in isolation, in clusters, or order into extended structures, such as vacancy planes.\cite{Ong2017,Wang2019} The former are typically very challenging to quantify using most imaging, spectroscopy, and scattering techniques, necessitating multi-modal and theory-backed characterization.\cite{Gunkel2020} Because even trace amounts of oxygen defects can significantly affect functionality, numerous studies have pursued better control of these important defects.\cite{Gunkel2020} Together, these and other defects give rise to distinct order-disorder behavior at oxide interfaces.

\section{Ordering During Synthesis}

Decades of study\cite{Huang2018,Brahlek2018} have shown that unexpected defects can dominate interface behavior, leading to a growing awareness that a simplistic picture of synthesis pathways and interface formation is insufficient. While less energetic ion species in MBE can help minimize intrinsic defect formation, the low growth pressure of the MBE process makes it difficult to control film stoichiometry.\cite{Stemmer2014} It has become increasingly apparent that thermodynamic and kinetic limitations, as well as slight fluctuations in process variables can greatly impact growth, which has spurred efforts to develop new synthesis approaches.\cite{MacManus-Driscoll2020} However, it is not trivial to achieve control over crystalline order in these highly non-equilibrium processes, due to the transient nature of the growth process and a lack of direct insight into the local growth environment, which together preclude effective predictive modeling. Nonetheless, there has been progress in controlling order-disorder behavior at interfaces, leveraging advanced \textit{in situ} methods, local probes, and increasingly intricate growth recipes. Here we highlight a selection of recent studies of perovskite- and pyrochlore-related materials, including Sr$_2$TiO$_4$ / SrTiO$_3$ (STO),\cite{Lee2014} LaTiO$_3$ (LTO) / STO,\cite{Cook2019} LaFeO$_3$ (LFO) / STO,\cite{Spurgeon2017,Comes2016a,Nakamura2016a,Nakamura2016} PbTiO$_3$ (PTO) / STO,\cite{Smith2017} BiFeO$_3$ (BFO) / STO,\cite{Smith2017} La$_{0.5}$Zr$_{0.5}$O$_{1.75}$ / LaAlO$_3$ (LAO),\cite{OSullivan2016} and Nd$_{0.5}$Zr$_{0.5}$O$_{1.75}$ / LAO,\cite{OSullivan2016} summarizing key observations.

\begin{figure*}
\includegraphics[width=\textwidth]{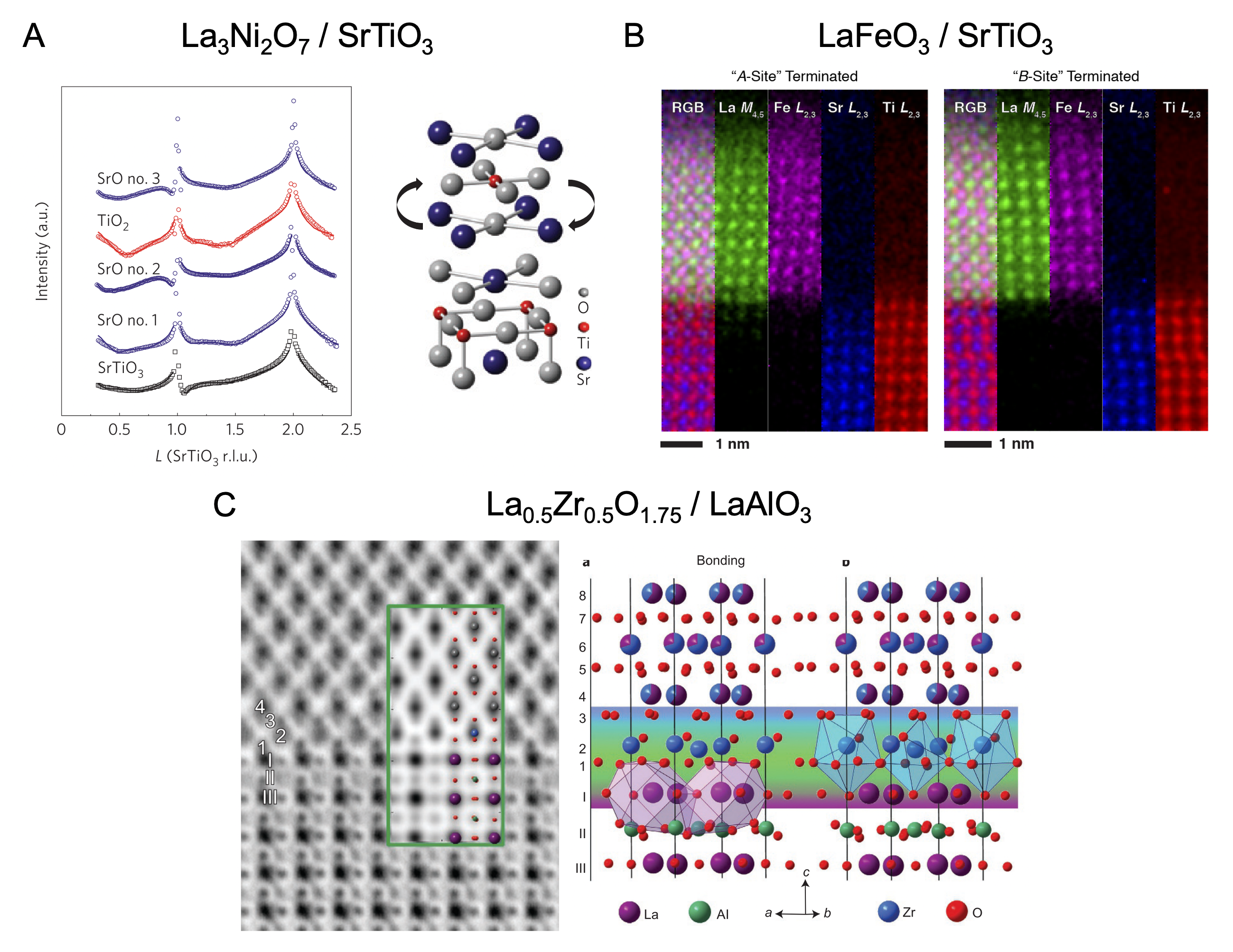}
\caption{Examples of complex synthesis pathways and dynamic rearrangement in interfaces of (A) La$_3$Ni$_2$O$_7$ / SrTiO$_3$. Reproduced from Reference \onlinecite{Lee2014} with permission of Springer Nature. (B) LaFeO$_3$ / SrTiO$_3$. Reproduced from Reference \onlinecite{Spurgeon2017} with permission of the American Physical Society. (C) La$_{0.5}$Zr$_{0.5}$O$_{1.75}$ / ﻿LaAlO$_3$. Reproduced from Reference \onlinecite{OSullivan2016} with permission of Spring Nature. \label{perovskite_growth}}
\end{figure*}

The study of STO perovskite heterointerfaces has consumed the oxides field for many decades. Much has been said about systems such as archetypal LAO / STO,\cite{Hwang2012, Ohtomo2004} which exhibits interface conductivity that is highly sensitive to interface configuration\cite{Nakagawa2006} and can be impacted by defects such as cation intermixing and oxygen vacancies.\cite{Chambers2010} The emergence of order on STO surfaces and dynamic interface rearrangement is a topic of intense discussion. Lee \textit{et al.}\cite{Lee2014} examined the layer-by-layer growth sequence of the Sr$_2$TiO$_4$ / STO system using a novel X-ray chamber equipped with an \textit{in situ} oxide MBE capability. They intentionally designed an SrO $\rightarrow$ SrO $\rightarrow$ TiO$_2$ $\rightarrow$ SrO growth sequence, observing changes in X-ray crystal truncation rod (CTR) patterns that reflect the film surface termination. Careful fitting of the CTR data showed that, while the preceding sequence was targeted, the as-synthesized interface rearranged into a SrO $\rightarrow$ TiO$_2$ $\rightarrow$ SrO configuration, as shown in Figure \ref{perovskite_growth}.A. They performed a series of \textit{ab initio} calculations to evaluate the energetics of layer formation and potential pathways for rearrangement. Interestingly, their results showed that there is a considerable thermodynamic driving force ($\sim 0.6$ eV / Ti atom) for structural rearrangement, which impedes stabilization of perovskite-derivative Ruddlesden-Popper (RP) phases. To explain this behavior, they performed molecular dynamics simulations, which revealed a potential pathway for rearrangement in which surface TiO$_2$ layers migrate into the underlying SrO layer through attack by TiO$_2$ + O tetrahedral clusters. Building off this knowledge, the authors deposited a SrO $\rightarrow$ SrO $\rightarrow$ SrO $\rightarrow$ TiO$_2$ sequence, which after rearrangement into SrO $\rightarrow$ SrO $\rightarrow$ TiO$_2$ $\rightarrow$ SrO allowed them to stabilize the desired RP phase. The authors obtained comparable results for a related La$_3$Ni$_2$O$_7$ structure, supporting the generalizability of this mechanism. Subsequent work by Nie \textit{et al.}\cite{Nie2014a} also showed that TiO$_2$ layers can migrate into underlying SrO layers during growth of titanate RP phases on DyScO$_3$ substrates, further supporting a complex and dynamic picture of film formation. These studies highlight the insufficiency of simple models for layer-by-layer growth and important insights provided by \textit{in situ} methods.

Dynamic layer rearrangement has also been observed in the LFO / STO system, where it affects the ability to control interfacial band structures for photocatalysis. Nakamura \textit{et al.}\cite{Nakamura2016} examined polar LFO grown on TiO$_2$- and SrO-terminated STO substrates using pulsed laser deposition (PLD), observing visible-light photocurrents that differed in direction depending on the interface configuration. They attributed this behavior to the differences in polarization that arise from interface dipoles and polar discontinuity. Subsequent work by Comes and Chambers\cite{Comes2016a} revealed that, despite attempts to engineer two interface polarities, the potential gradient of the LFO and interface band offsets are similar. This contradictory finding motivated further investigation by Spurgeon \textit{et al.}\cite{Spurgeon2017} to determine mechanisms for the observed behavior. They examined the two interface configurations, with initial STO terminations confirmed to be TiO$_2$ and SrO by X-ray photoelectron spectroscopy (XPS) prior to LFO deposition. After the LFO layer was deposited, STEM-EELS was used to assess the character of the resulting interfaces. As shown in Figure \ref{perovskite_growth}.B, the final heterojunctions were identical, with LaO $\rightarrow$ TiO$_2$ configurations in both cases. While EELS fine structure measurements revealed a subtle shift in Fe valence for the TiO$_2$-terminated sample, there was no clear long range intermixing, indicating that the SrO-terminated sample undergoes a structural rearrangement similar to the work of Lee \textit{et al.}\cite{Lee2014} The authors conducted \textit{ab initio} calculations to assess possible drivers for rearrangement, finding that an FeO$_2$ $\rightarrow$ SrO configuration is much less energetically preferred than the LaO $\rightarrow$ TiO$_2$ configuration. Moreover, they found that during sequential, shuttered growth, unstable Fe$^{4+}$ ions can form, which in turn can promote oxygen vacancies. These vacancies may act as a pathway for the rearrangement to occur. These findings are aligned with \textit{in situ} XRD studies performed by Cook \textit{et al.}\cite{Cook2019} during PLD growth of LTO / STO. They observed the presence of a TiO$_2$ double layer at STO substrate surface that persisted through continuous interface rearrangement, regardless of the LTO film growth conditions. This series of studies shows how dynamic crystal rearrangements mediate the emergence of order at interfaces. A multimodal approach based on local probes and atomistic simulations can help map energetic pathways.

With these limitations in mind, other studies have attempted to balance thermodynamic and kinetic factors to more precisely control growth. Smith \textit{et al.}\cite{Smith2017} grew PTO and BFO on STO substrates using both continuous codeposition and adsorption-controlled growth modes in MBE. The resulting films were examined using XRD to determine the relative distribution of phases and their deviation from the targeted, phase-pure structures. They observed that phase-pure PTO can only be achieved during continuous codeposition with low titanium flux. The growth of this system cannot be described by thermodynamics alone, since it is strongly dependent on titanium flux rate, meaning it is kinetically driven. The authors obtained similar results for BFO growth, finding that the selection of oxidant mixture can impact the growth window for phase-pure BFO. Efforts to synthesize pyrochlore-like structures has revealed similar effects. O'Sullivan \textit{et al.}\cite{OSullivan2016} examined deposition of the pyrochlore-derived defect fluorites La$_{0.5}$Zr$_{0.5}$O$_{1.75}$ and Nd$_{0.5}$Zr$_{0.5}$O$_{1.75}$ on LAO substrates using PLD. The films were measured using XRD, as well as local STEM-EELS and annular bright field (ABF) imaging, the latter of which revealed a complex interfacial reconstruction to accommodate mismatch between the perovskite and defective fluorite layers. As shown in Figure \ref{perovskite_growth}.C, the resulting interface is not a simple combination of two structures, but rather a distinct region containing both structural and chemical distortions. The authors performed \textit{ab initio} calculations for various possible interfacial rearrangements, which they then fitted to the ABF data. They determined that the ability of the La$_{0.5}$Zr$_{0.5}$O$_{1.75}$ fluorites to accommodate eight-fold coordinated fluorite, six-fold coordinated perovskite $B$-sites, and intermediate seven-fold coordinated perovskite-fluorite hybrid blocks enables coherent matching of the film and substrate. An important conclusion of this work was that, depending on the coordination chemistry of the materials under consideration, it is possible to harness structural reorganization to coherently interface drastically different interface components.

This selection of recent work underscores the complexities and rapidly evolving picture of layer-by-layer materials synthesis and the emergence of order in thin film interfaces. It is increasingly apparent that simplistic models of film growth can overlook kinetic and thermodynamic limitations, which in turn drive complex structural rearrangements that lead to deviations from targeted structures. The contributions of \textit{in situ} measurement techniques and local probes are invaluable in visualizing these processes, and more development is needed to map and harness synthesis pathways to achieve desired structures.

\section{Disorder During Irradiation}

The interaction of radiation with matter is a complex process with implications for many fields, including electrochemistry, photonics, and nuclear energy. Oxides have long been studied in this context, since they comprise many of the materials used in energy storage, the nuclear fuel cycle, and devices in extreme environments. A large body of work\cite{Lang2010,Ewing2004,Wuensch2000,Ewing1995} has examined bulk pyrochlore oxides as potential nuclear waste forms and sensors, which possess unique radiation tolerance because of the structure's ability to disorder into a defected fluorite structure.\cite{Sickafus2000} Perovskite materials have also been studied for these applications \cite{Zhang2009,Zhang2006,Ewing1995} and both systems have attracted attention as media for atom-by-atom materials synthesis using electron beam irradiation.\cite{Sachan2018,Jesse2016a,Jesse2015}

While radiation-induced disorder has been studied in both bulk pyrochlores and perovskites, less attention has been paid to model interfaces of these and other oxides. Recent reviews\cite{Zhang2018a,Beyerlein2015,Beyerlein2013} have emphasized the unique environment of nanostructured oxides, their potential enhanced radiation tolerance, and a lack of literature concerning structured interface responses. A handful of oxide interfaces have been examined in detail using modern local characterization methods, yielding a wide variety of different behaviors. Here we review a selection of recently examined perovskite systems, including TiO$_2$ / STO,\cite{Zhuo2012,Zhuo2011} CeO$_2$ / STO,\cite{Dholabhai2014,Aguiar2014} MgO / STO,\cite{Aguiar2014a}  and BaTiO$_3$ (BTO) / STO,\cite{Bi2013} as well as pyrochlore systems, including Gd$_2$Ti$_2$O$_7$ (GTO) / Y:ZrO$_2$,\cite{Kreller2019,Janish2020} La$_2$Ti$_{2-x}$Zr$_x$O$_7$ (LTZO) / Y:ZrO$_2$,\cite{Sassi2019,Kaspar2017} and La$_2$Ti$_2$O$_7$ / SrTiO$_3$ (STO).\cite{Spurgeon2020,Kaspar2018} We identify overall trends and suggest directions for future study.

\begin{figure*}
\includegraphics[width=0.8\textwidth]{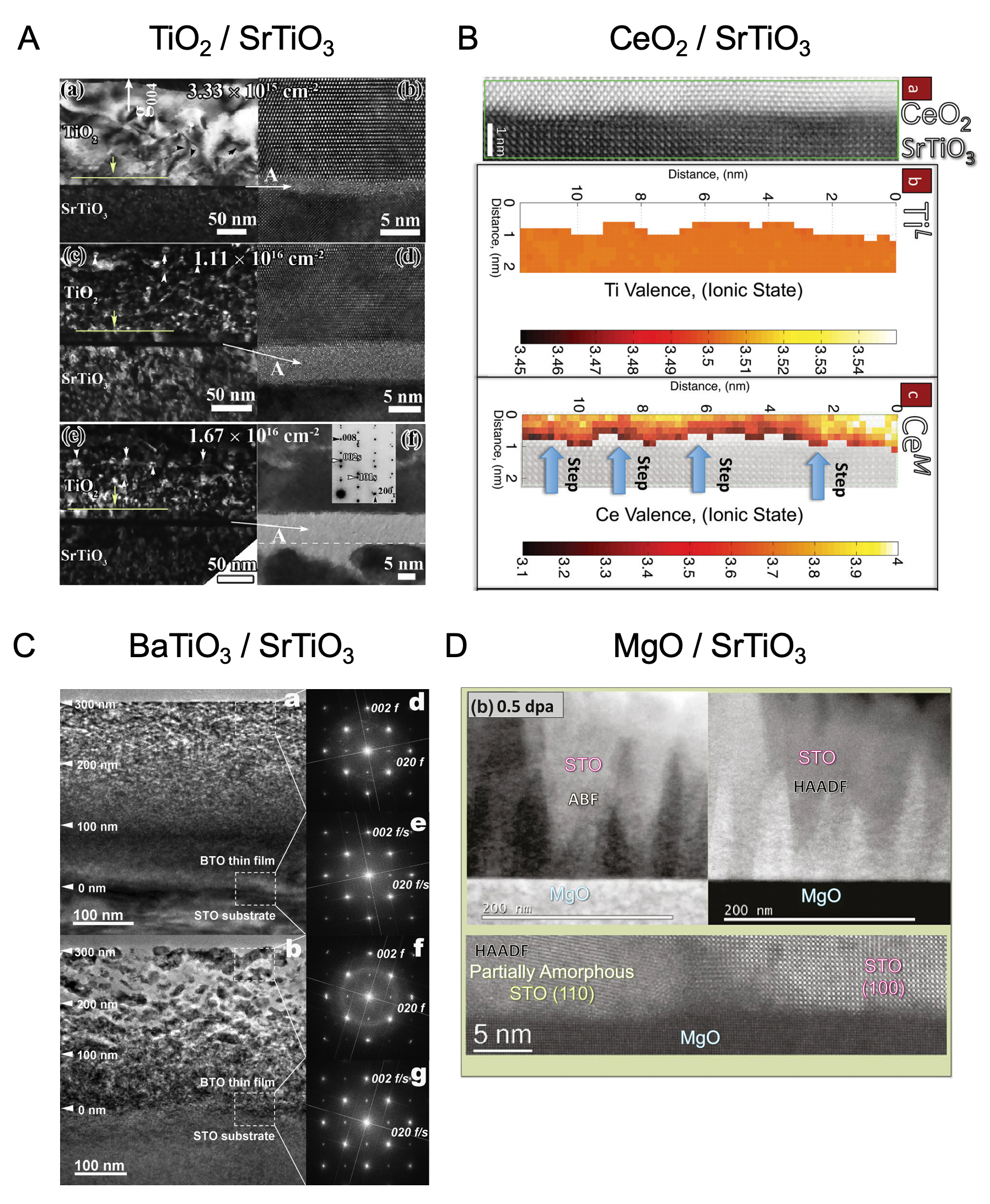}
\caption{Examples of irradiation-induced disorder in perovskite interfaces of (A)  TiO$_2$ / SrTiO$_3$. Reproduced from Reference \onlinecite{Zhuo2011} with permission of Elsevier. (B) CeO$_2$ / SrTiO$_3$. Reproduced from Reference \onlinecite{Aguiar2014} with permission of Wiley. (C) BaTiO$_3$ / SrTiO$_3$. Reproduced from Reference \onlinecite{Bi2013} with permission of the American Institute of Physics. (D) MgO / SrTiO$_3$. Reproduced from Reference \onlinecite{Aguiar2014a} with permission of the Materials Research Society. \label{perovskite_irrad}}
\end{figure*}

A central question in the study of radiation effects in oxide heterojunctions concerns the potentially unique response of the interface relative to the bulk of the component materials. Ubiquitous STO interfaces, which can be synthesized with a high degree of control, have unsurprisingly attracted considerable study to help answer this question. As shown in Figure \ref{perovskite_irrad}.A, Zhuo \textit{et al.}\cite{Zhuo2012,Zhuo2011} have examined STO / TiO$_2$ interfaces irradiated with 250 keV Ne$^{2+}$ ions using a cross-sectional TEM and atomistic simulations. The authors observed a defect-denuded zone extending a nearly constant $\sim 10$ nm on the TiO$_2$ side of the interface, as well as an amorphous layer on the STO side that grows with increasing dose. Prior to irradiation, the authors noted a high density of misfit dislocations on the TiO$_2$ side, which are known pathways for defect annihilation.\cite{Kolluri2010} However, even after subsequent irradiation and disappearance of the dislocations, the denuded zone remained, pointing to an alternative mechanism. Atomistic modeling indicated that the interface environment itself appears to have little effect on defect migration and it is rather competition between the chemical potential and kinetics of the two interface components that drives defect dynamics. Moreover, the greater susceptibility of the STO to radiation damage versus TiO$_2$ appears to dictate which side of the interface is prone to disorder.

The question of interface contributions to disordering of perovskite-based systems has also been examined in the STO / CeO$_2$ system by Aguiar \textit{et al.},\cite{Aguiar2014} who focused on the role of STO substrate step structures. As shown in Figure \ref{perovskite_irrad}.B, the authors used STEM-EELS to probe subtle local perturbations in the chemistry of the adjacent CeO$_2$ layer around step edges in the starting material. They observed a Ce valence shift toward Ce$^{3+}$ at the interface that was even more pronounced around TiO$_2$ than SrO terrace steps. During irradiation by 400 keV Ne$^{2+}$ ions, they measured amorphization of the STO side similar to the STO / TiO$_2$ interface, but found that damage tended to cluster around step edges and was associated with a reversion to the stoichiometric Ce$^{4+}$ valence on the CeO$_2$ side. These findings led the authors to propose that sub-stoichiometric CeO$_2$ regions may act as sinks for radiation-induced oxygen interstitials, which migrate from the STO side, leaving behind structurally destabilizing oxygen vacancies. Dholabhai \textit{et al.}\cite{Dholabhai2014} used atomistic simulations to study the response of cation and anion defects for different step edge structures in the same system. They found that certain step structures may act as recombination sites for point defects, but determined that the recombination process is also significantly impacted by defect mobility. In particular, O vacancies and interstitials are more mobile in STO than their Sr and Ti counterparts, leading to the annihilation of anion defects and accumulation of cation defects that destabilize the STO faster than CeO$_2$.

The above findings reinforce the competition between interface configuration and defect kinetics in interface components during disordering. For example, in the case of STO / BTO irradiated by 300 keV Ne$^{2+}$, Bi \textit{et al.}\cite{Bi2013} found no evidence for interface denuded zones or preferential amorphization. While they did observe a relationship between lattice strain state and radiation tolerance, their atomistic simulations showed no preference for defect formation on either side of the interface. Rather, they found that subtle differences in defect mobilities can impact damage recovery and its spatial onset relative to the interface position. These results are supported by numerical simulations performed by Blas \textit{et al.},\cite{Uberuaga2018,PedroUberuaga2013} which have shown that defect formation tendencies and transport in interface components, rather than the unique response of the interface itself, may determine overall response. In contrast, Aguiar \textit{et al.}\cite{Aguiar2014a} have shown that the behavior of STO / MgO is strongly influenced by interface configuration. They synthesized a structure consisting of (100)- and (110)-oriented STO domains on MgO, which were subsequently irradiated using 250 keV Ne$^{2+}$ ions. They observed the preservation of a crystalline denuded zone in the STO (100) domains but not the STO (110) ones, which they attributed to differences in the charge state of the two orientations. Charge neutral STO (100) planes are less likely to interact strongly with the space charge induced by irradiation than STO (110) planes, favoring the preservation of the former over the latter. In this instance, the exact configuration of the interface appears to play a dominant role in the disordering process.

\begin{figure*}
\includegraphics[width=0.8\textwidth]{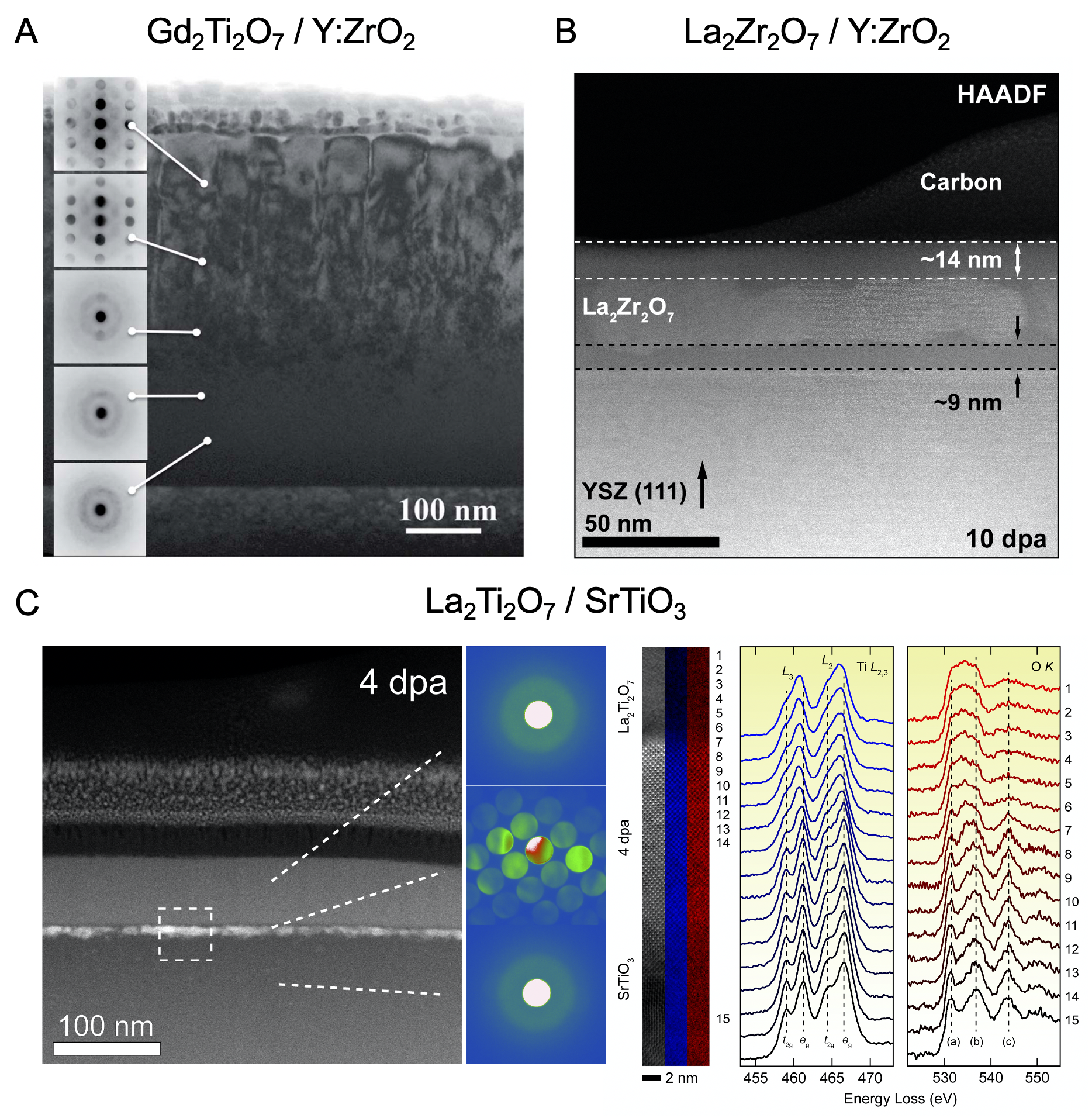}
\caption{Examples of irradiation-induced disorder in pyrochlore interfaces of (A) Gd$_2$Ti$_2$O$_7$ / Y:ZrO$_2$. Reproduced from Reference \onlinecite{Kreller2019} with permission of the Royal Society of Chemistry. (B) La$_2$Zr$_2$O$_7$ / Y:ZrO$_2$. Reproduced from Reference \onlinecite{Kaspar2017} with permission of Elsevier. (C) La$_2$Ti$_2$O$_7$ / SrTiO$_3$. Reproduced from Reference \onlinecite{Spurgeon2020} under CC-BY-4.0 license.\label{pyrochlore_irrad}}
\end{figure*}

As already mentioned, radiation damage effects have been extensively studied in bulk pyrochlores, but fewer efforts have focused on well-controlled thin film interfaces of these materials. As shown in Figure \ref{pyrochlore_irrad}.A, Kreller \textit{et al.}\cite{Kreller2019} have examined GTO / YSZ irradiated by 200 keV He$^+$, utilizing TEM and selected-area diffraction to locally visualize the disordering process. They observed graded structural amorphization consistent with the expected ion irradiation damage profile and no clear differences at the film-substrate interface. They did note abrupt and substantial changes in ionic conductivity during irradiation, which were linked to underlying anion disorder and potentially impacted by substrate-imparted lattice strain. Importantly, these changes in conductivity appear to be solely related to structural disorder, underscoring their role in determining material behavior.

Interface effects on the disordering behavior of pyrochlores have also been explored in the LTZO system. Kaspar \textit{et al.}\cite{Kaspar2017} studied La$_2$Zr$_2$O$_7$ (LZO) / YSZ (111) interfaces irradiated by 1 MeV Zr$^+$ ions. The authors examined the stages of disordering using area-averaged scattering techniques and local high-resolution STEM imaging. Their measurements revealed a very unusual progression of disorder in which both the film surface and film-substrate interface preferentially amorphized relative to the bulk of the LZO. Geometric phase analysis (GPA) measurements revealed a network of misfit dislocations, whose surrounding strain fields are known to affect local structure, chemistry, and radiation response.\cite{Shutthanandan2017,Choudhury2014} Structural relaxation may also partially account for the observed behavior at the film surface. To gain further examine disordering dynamics in this system, Sassi \textit{et al.}\cite{Sassi2019} conducted an \textit{ab initio} molecular dynamics study of electronic excitations in various LTZO compounds. While they did not specifically focus on interface behavior, their simulations elucidated potential mechanisms governing disordering processes. They found that the formation of O$_2$-like molecules mediates amorphization and that, depending on the phase of the material, TiO$_6$ octahedral rotations can promote this formation. Doping of Zr into the monoclinic La$_2$Ti$_2$O$_7$ phase increases the material's bandgap and mitigates these rotations, making it more resistant to electronic excitations. These results point to the potentially important role of interfaces in constraining these octahedral distortions, analogous to well-known substrate clamping in perovskites.\cite{Nord2019}

To better understand the nature of electronic defects and the role of interface configurations, Spurgeon \textit{et al.}\cite{Spurgeon2020} examined La$_2$Ti$_2$O$_7$ / STO interfaces irradiated by 1 MeV Zr$^+$ ions. The authors employed analytical STEM imaging and nanobeam diffraction to visualize changes in local crystallinity imparted by irradiation. They observed a progression of amorphization, beginning with the surface of the film and then the bulk of the substrate, with the preservation of a distinct crystalline band on the STO side of the interface. STEM-EELS measurements of the as-grown and irradiated films showed substantial changes in the O $K$ and Ti $L_{2,3}$ edge fine structure, reflecting the formation of oxygen vacancies. STEM imaging also revealed two predominant interface configurations---a majority monoclinic La$_2$Ti$_2$O$_7$ / STO and minority perovskite LaTiO$_3$ / STO. To determine possible drivers for disordering behavior, the authors conducted \textit{ab initio} simulations to calculate the energy of vacancy formation for the two configurations. Their results showed that oxygen vacancies are difficult to form at the interface, with a distinct tendency for vacancies to accumulate on the film side in the former and in the substrate in the latter. These results suggest that that relative susceptibility of the two interface components can perhaps dictate the formation and migration of oxygen vacancies and other radiation-induced defects. Moreover, this result is similar to the work of Aguiar \textit{et al.}\cite{Aguiar2014a} on STO / MgO, which showed a preference for denuded zone formation in STO (001) versus STO (110) domains. It suggests that appropriate selection of phases, chemistries, and interface configurations may be used to guide the radiation response of oxide heterostructures.

The highlighted perovskite and pyrochlore studies illustrate the complex nature of radiation-induced disorder at oxide interfaces. Emerging local probes and modeling efforts have uncovered a wide variety of interface behaviors that are intimately connected to the ability to form and transport point defects. Still, more systematic studies of salient factors governing the radiation response are needed to isolate interface and bulk effects on defect dynamics.

\section{Conclusions and Future Directions}

In this critical review, we have attempted to summarize a selection of recent developments in understanding the origins of of order-disorder behavior during the synthesis and irradiation of oxide interfaces. Great strides have been made in the use of \textit{in situ} methods, local chemical imaging, and predictive modeling to probe the evolution of many important oxide systems. Nonetheless, our survey shows that more work is needed to disentangle and harness the complex energetic drivers for order-disorder behavior.

At present we are limited in our ability to visualize and model highly non-equilibrium processes during synthesis and disordering. \textit{In situ} X-ray scattering methods can provide valuable insight into dynamic crystal rearrangements during growth. These methods should be extended to other systems, as well as broader kinetic and thermodynamic parameter spaces. Combinatorial synthesis approaches\cite{Bollinger2016} and data science techniques\cite{Provence2020a} may help explore many possible synthesis pathways and extract the most optimal ones. In addition, \textit{in situ} studies using other techniques such as STEM should also become more commonplace. Janish \textit{et al.},\cite{Janish2020} for example, have shown that \textit{in situ} STEM heating experiments can visualize the recovery of radiation damage in pyrochlores. This new experimental window will in turn help refine key inputs for \textit{ab initio} models of order-disorder behavior, guiding targeted synthesis efforts.

In regards to radiation effects, more systematic study of the factors that mediate defect formation and transport is needed. Are substrate-induced strain and interface configuration important or is it simply the competition of defect formation in each component that matters? Can heterostructuring be used to craft composite oxide structures with tunable radiation responses, as has been shown in the case of metal multilayers?\cite{Zhang2018a} And how can defect engineering be used to direct disordering pathways? These complicated questions will require careful study of well-defined model oxide systems under controlled irradiation conditions. Ultimately, a better understanding of order-disorder processes will allow us to more fully harness this important class of functional materials.

\section{Acknowledgements}

I would like to thank Drs. Matthew Olszta and Michel Sassi for reviewing this manuscript. This research was supported by the Laboratory Directed Research and Development (LDRD) Nuclear Processing Science Initiative (NPSI) at Pacific Northwest National Laboratory (PNNL). PNNL is a multiprogram national laboratory operated for the U.S. Department of Energy (DOE) by Battelle Memorial Institute under Contract No. DE-AC05-76RL0-1830. A portion of the STEM imaging shown was performed in the Radiological Microscopy Suite (RMS), located in the Radiochemical Processing Laboratory (RPL) at PNNL.

\clearpage

\makeatletter
\renewcommand{\@biblabel}[1]{[#1] }
\makeatother
\bibliography{references}

\end{document}